\documentclass[conference]{IEEEtran}
\IEEEoverridecommandlockouts

\usepackage{cite}
\usepackage{amsmath,amssymb,amsfonts}
\usepackage{graphicx}
\usepackage{textcomp}

\usepackage{stfloats}
\usepackage{url}
\usepackage{verbatim}
\usepackage{graphicx}
\usepackage{cite}

\usepackage{algorithm}
\usepackage{algpseudocode}

\usepackage{array}
\usepackage{cleveref}

\usepackage[table, x11names]{xcolor}

\usepackage[caption=false,font=normalsize,labelfont=sf,textfont=sf]{subfig}

\def\BibTeX{{\rm B\kern-.05em{\sc i\kern-.025em b}\kern-.08em
    T\kern-.1667em\lower.7ex\hbox{E}\kern-.125emX}}
\begin{document}

\bstctlcite{IEEEexample:BSTcontrol}	

\renewcommand{\IEEElabelindentfactori}{0}

\title{Reactive Orchestration for Hierarchical Federated Learning Under a Communication Cost Budget}

\author{
    \IEEEauthorblockN{Ivan Čilić\IEEEauthorrefmark{1}, 
    Anna Lackinger\IEEEauthorrefmark{3}, 
    Pantelis A. Frangoudis\IEEEauthorrefmark{3}, 
    Ivana Podnar Žarko\IEEEauthorrefmark{1}, 
    Alireza Furutanpey\IEEEauthorrefmark{3}, \\
    Ilir Murturi\IEEEauthorrefmark{3}, 
    Schahram Dustdar\IEEEauthorrefmark{3}} \\\vspace{-3mm}
    \IEEEauthorblockA{\IEEEauthorrefmark{1}Faculty of Electrical Engineering and Computing, University of Zagreb, Zagreb, Croatia
    \\\{ivan.cilic, ivana.podnar\}@fer.hr} 
    \IEEEauthorblockA{\IEEEauthorrefmark{3}Distributed Systems Group, TU Wien, Vienna, Austria
    \\\{a.lackinger, p.frangoudis, a.furutanpey, imurturi, dustdar\}@dsg.tuwien.ac.at}
    \vspace{-1cm}
}

\maketitle

\begin{abstract}
Deploying a Hierarchical Federated Learning (HFL) pipeline across the computing continuum (CC) requires careful organization of participants into a hierarchical structure with intermediate aggregation nodes between FL clients and the global FL server. This is challenging to achieve due to (i) cost constraints, (ii) varying data distributions, and (iii) the volatile operating environment of the CC. In response to these challenges, we present a framework for the adaptive orchestration of HFL pipelines, designed to be reactive to client churn and infrastructure-level events, while balancing communication cost and ML model accuracy. Our mechanisms identify and react to events that cause HFL reconfiguration actions at runtime, building on multi-level monitoring information (model accuracy, resource availability, resource cost). Moreover, our framework introduces a generic methodology for estimating reconfiguration costs to continuously re-evaluate the quality of adaptation actions, while being extensible to optimize for various HFL performance criteria. By extending the Kubernetes ecosystem, our framework demonstrates the ability to react promptly and effectively to changes in the operating environment, making the best of the available communication cost budget and effectively balancing costs and ML performance at runtime.
\end{abstract}


\section{Introduction} 
\label{intro}

Federated learning (FL) trains machine learning (ML) models on a federation of (edge) devices without the need to centrally store and process the raw data on cloud servers~\cite{mcmahan}. Data remains on (IoT or other edge) devices, called \emph{clients}, and local ML model versions are sent to a central aggregator typically running in the cloud. The aggregator creates a global model which is sent back to the FL clients for the subsequent training round. This iterative process continues until a convergence criterion is met. FL thus improves privacy and significantly reduces the computational and storage costs of the central server.

FL faces a number of challenges~\cite{kairouz} due to the \textit{heterogeneity} of the environments in which it operates.
Clients typically run on (i) different hardware which is often resource-constrained and unreliable, (ii) use different communication protocols to exchange ML models with the aggregator, or (iii) train on data that are unbalanced and often not independent and identically distributed (non-IID). Hardware heterogeneity delivers different training performances and leads to the occurrence of slow clients, i.e., stragglers~\cite{Reisizadeh}. Furthermore, although raw data do not have to be transported, the exchange of model updates can be costly, particularly for large-scale FL tasks and when training large models.
In IoT use cases where FL is highly relevant~\cite{Nguyen21fliot}, this problem is even more severe as devices often operate over unstable and bandwidth-limited networks.

Hierarchical federated learning (HFL)~\cite{liu_hiear} emerged in response to these challenges. HFL places intermediate nodes at the edge of the computing continuum (CC), closer to FL clients, to perform local aggregation before sending the aggregated models to the global aggregator. It has been shown that this has the potential to reduce communication costs and overall training time~\cite{liu_hiear, Lin24DFL}, thus also saving energy~\cite{Rango23}.

HFL pipelines are well suited for execution in the CC, as their multi-layered architecture fits seamlessly into the decentralized and tiered structure of the CC. However, configuring an HFL pipeline is not trivial: Aggregator nodes need to be strategically placed in the CC so that clients are assigned to them considering not only resource-level information and network distance, but also ideally taking into account the characteristics of individual client datasets~\cite{deng_hier}. These most likely vary in volume and contain non-IID data, which almost inevitably leads to performance degradation of an FL task~\cite{zhu_noniid}. Importantly, the volatility of the CC suggests that, during the lifetime of an HFL pipeline, changes in the operating environment will occur and impact it, potentially severely, e.g., by degrading model accuracy or generating considerable network traffic. Therefore, the HFL pipeline must be continuously reconfigured during its runtime. 

While the literature on HFL is extensive (\S\,\ref{section:rel}), system support and mechanisms for continuous HFL adaptations are rather understudied. We bridge this gap by introducing \emph{a framework for adaptive and continuous orchestration of HFL pipelines} tailored to dynamic CC environments. 
In particular, we make the following contributions:

\noindent\textbf{HFL Orchestrator~(\S\,\ref{sec:orchestration}).} We introduce an HFL orchestrator that is extensible to support various configuration strategies, such as communication cost minimization, balanced data distribution, or combinations thereof. Our special-purpose orchestrator interfaces with general-purpose orchestrators (Kubernetes and its edge-centric variants) to deploy the entities of the HFL pipeline on CC hosts, monitors the execution of the pipeline at the resource and service levels, and reconfigures the pipeline at runtime if needed. Our orchestration framework is available as open source.\footnote{\url{https://github.com/AIoTwin/fl-orchestrator/tree/icmlcn}}

\noindent\textbf{Algorithm~(\S\,\ref{section:reconf}).} Reconfiguration actions can take place on various dimensions (e.g., attach a client to a different aggregator node, adapt the model aggregation frequency), and involve a direct cost (e.g., cost to transfer container images and initialize a new aggregator node) and a post-reconfiguration impact that could be positive or negative. We introduce a generic methodology to model these costs and adopt a hybrid reactive-predictive orchestration approach: The HFL orchestrator \emph{reacts} to a relevant event (e.g., a new client joins) by computing and applying a new configuration, and executes a Reconfiguration Validation Algorithm (RVA) that continuously evaluates reconfiguration decisions by \emph{predicting} their future impact in terms of operational costs (particularly, communication-related) and model accuracy, \emph{reverting} them upon a negative performance forecast. 

\noindent\textbf{Evaluation~(\S\,\ref{section:eval}).} We evaluate our approach on HFL tasks using the CIFAR-10 dataset~\cite{cifar-10} over a testbed including a K3s cluster with commodity hardware. Our results demonstrate that our scheme promptly reacts to environmental changes such as client churn, effectively balancing communication costs and ML performance during runtime.

\section{Hierarchical Federated Learning (HFL) Orchestration}
\label{sec:orchestration}

\subsection{Preliminaries}
\label{sec:preliminaries}
\begin{figure}[htbp]
\centering
\includegraphics[width=3.4in]{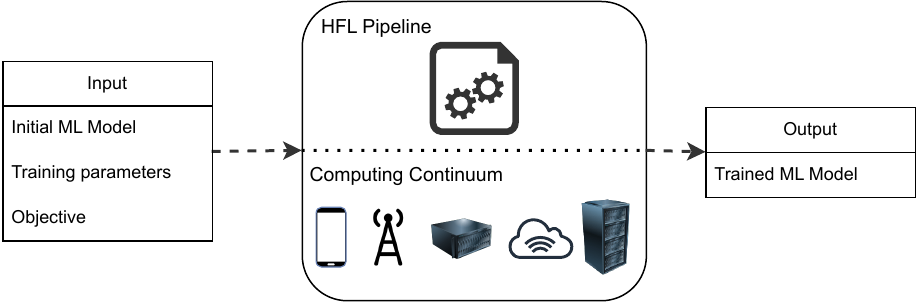}
\caption{HFL task definition.}
\label{fig:hfl-task}
\end{figure}

Contrary to traditional FL where there is a single global aggregator (GA) typically residing in the cloud, HFL introduces \emph{local aggregators (LAs)} at the edge.
In a typical HFL setup, an ML engineer defines an HFL task that runs within an HFL pipeline in the CC, as outlined in \figurename~\ref{fig:hfl-task}. The HFL task is defined with the following inputs: (i) an initial ML model, (ii) training parameters, and (iii) the \emph{orchestration objective}. The initial ML model serves as the starting point from which all clients begin their training during the initialization phase of the pipeline. The training parameters include batch size, learning rate, etc. The orchestration objective can be defined on a case-by-case basis, for example, to either optimize the model's performance within a predefined communication cost budget or to minimize the cost while achieving a specific performance target, such as a target level of accuracy or loss. Other objectives are also possible and can be supported by an HFL orchestrator.
A typical HFL pipeline is executed through the following phases: 
\begin{enumerate}
    \item \textbf{Initialization phase}: Clients prepare their local data for training and the initial ML model is distributed by the GA to all clusters and, in turn, to the clients.
    \item \textbf{Local training}: Clients train their local models for a predefined number of epochs.
    \item \textbf{Local aggregation}: After local training, clients send their model updates to their LA for aggregation. The aggregated model is distributed back to the clients in the cluster.
    \item \textbf{Global aggregation}: After a predefined number of local aggregations, usually called \textit{local rounds}, the local model is sent to the GA for global aggregation. Subsequently, the global model is distributed back to the LAs which forward it to the clients in their cluster. This marks the beginning of a new global round.
    \item \textbf{Model convergence and completion}: When the model has converged, or a certain threshold has been reached, the pipeline execution is completed.
\end{enumerate}

\subsection{HFL pipeline configuration}
An HFL pipeline is characterized by its \emph{configuration}, which is defined to contain the following elements.

\begin{itemize}
    \item \emph{Topology}. The topology of an HFL pipeline defines which nodes residing in the CC are used by the pipeline, which roles these nodes take and how the nodes are interconnected based on the roles. 
    Compared to traditional flat FL, the topology in a hierarchical FL setup is extended to define node clusters and the placement of LAs at the edge and the GA in the cloud. A \emph{client selection} process can take place before the initialization phase and between local or global training rounds~\cite{fu} and, compared to flat FL, should also consider the dimension of client-LA association. Topology formation criteria can account for node resources, network connectivity, or per-client data volume.
    
    \item \emph{Aggregation algorithm}. A configuration includes the selected aggregation algorithm (e.g., FedAvg \cite{mcmahan}, FedAvgM \cite{fedavgm}, or FedOpt \cite{fedopt}), which determines how model updates are aggregated at the local and global level. These algorithms can be classified into two categories depending on how they perform the aggregation: synchronously or asynchronously. In this paper, we only consider synchronous aggregation.
    \item \emph{Aggregation frequency}. It is defined as the number of local iterations before a model is sent for aggregation. In HFL, two parameters define this frequency: (i) \textit{local epochs}, which are used by clients to create local models that are sent for local aggregation to the LAs, and (ii) \textit{local rounds} which specify the number of local model update-aggregation rounds performed by the LAs before the models are sent to the GA. Aggregation frequency has a direct impact on communication costs. In addition, the experiments reported in~\cite{deng_hier} demonstrate that the frequency of model aggregation can significantly affect FL performance when distributed nodes have imbalanced training data.
\end{itemize}

Changes in the operating environment of an HFL pipeline, such as nodes joining and leaving, node resource fluctuations, or network disruptions can have various effects on pipeline execution.
For example, a client becoming unavailable can introduce degradation of the ML model performance, or a network change can increase the communication cost of the pipeline and this can faster deplete a potential cost budget defined for a task. Responding to such events requires an adaptive orchestration mechanism that maintains the HFL pipeline deployed with the best HFL configuration for the given state of the CC, i.e., events cause pipeline reconfigurations. 

Carrying out the appropriate \emph{reconfigurations} is however non-trivial, as applying a new configuration at runtime is not guaranteed to provide better ML model performance in the future. 
Predicting the impact of a reconfiguration (e.g., re-assigning client nodes to LAs) before applying it would be invaluable for an HFL orchestrator. To this end, one could reasonably apply approaches that quantify and account for node utility~\cite{fu}. However, our pilot experiments have shown that a node's utility does not always reflect on the ML model performance as expected. Instead, motivated by our experimental findings, we implement reconfiguration validation \emph{reactively}: Our HFL orchestrator computes and applies a new configuration it deems optimal, i.e., the best-fit configuration, monitors its short-term effects, and uses this information to forecast whether it should maintain the new configuration or revert to the previous one, while taking into account the specific orchestration objective.

\subsection{HFL orchestrator design}
\label{section:arch}
We adopt a two-level orchestration design, making a distinction between HFL service-specific and general-purpose orchestration tasks. Our HFL orchestrator makes service-specific pipeline configuration decisions, which translates to actionable input for a General Purpose Orchestrator (GPO) -- in our case Kubernetes or its edge-centric derivatives. Our orchestration framework carries out the following functions:
\begin{enumerate}
    \item deploying and managing the entities of the HFL pipeline, i.e., clients and aggregators;
    \item monitoring the execution of the pipeline at the infrastructure and HFL service levels; and
    \item performing HFL pipeline reconfiguration as needed. 
\end{enumerate}

Our framework collects infrastructure characteristics by interfacing with the GPO, as well as monitors HFL-specific information (e.g., historical learning performance information, whether a node hosts data available for training and can thus assume the role of a client or it can only be used for aggregation, etc.). Furthermore, it allows the user to define the orchestration objective, which serves as the optimization goal for the orchestrator (e.g., to save on communication cost, to minimize training time, etc.). In this paper, we focus on the following objective: \emph{Providing the best ML model performance under a user-specified total available communication cost budget $B$}. Note that our orchestrator also supports different objectives, such as minimizing total cost to reach a user-specified target model accuracy. Our modular design allows to incorporate and activate on demand existing state-of-the-art configuration strategies to meet an orchestration objective, such as the ones proposed in \cite{deng_hier, abdellatif_hier, wang_hier}. 

Once a configuration is derived -- either at pipeline deployment time or in response to events -- the GPO is invoked to deploy containerized versions of the HFL entities (termed \emph{HFL service instances}) over CC hosts. An HFL service instance can be either a client or an aggregator. This service runs with a sidecar HFL agent that is designed to monitor the progress of an HFL task and send reports to the HFL orchestrator component. In the interest of space, we omit further technical details of our orchestration framework.

\section{HFL pipeline reconfiguration}
\label{section:reconf}

HFL pipeline reconfiguration is triggered by various events monitored by our orchestrator. These events can be categorized as either infrastructure-related or related to ML model performance metrics. Reconfiguration is triggered reactively when an event such as a node failure or a significant increase in training loss occurs. 
Reconfiguration does not come for free: The calculation of a new pipeline configuration introduces the computational cost of solving a complex optimization problem, while additional HFL service artifacts may have to be deployed, interrupting the FL flow. Reconfiguration costs are particularly increased in environments with a high number of clients and potential aggregation nodes. Nevertheless, pipeline reconfiguration may enhance HFL performance to such an extent that the invested cost is fully recouped.

\begin{figure}[htbp]
\centering
\includegraphics[width=3.4in]{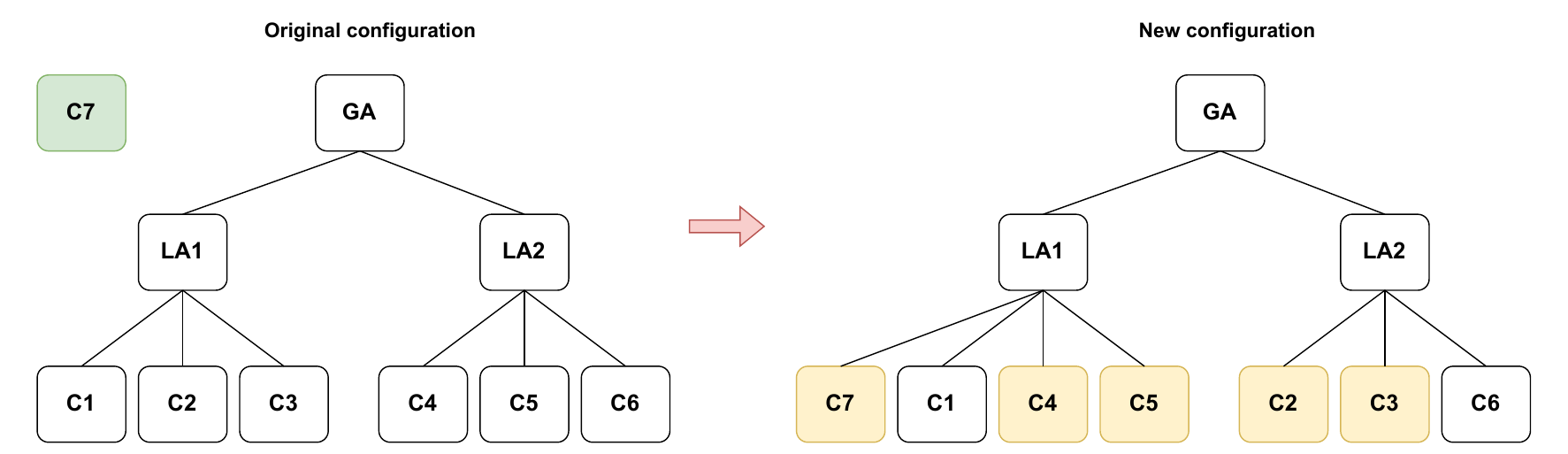}
\caption{Reconfiguration trigger: a new client joins the pipeline.}
\label{fig:reconf-changes}
\vspace{-2mm}
\end{figure}

Whatever the trigger, reconfiguration is performed in the same manner. First, a new best-fit configuration has to be identified considering the new environmental conditions. Next, the HFL orchestrator checks the differences between the \textit{original configuration}, i.e., the one before reconfiguration, and the \textit{new configuration}, i.e., the one after reconfiguration, to define changes necessary to apply the reconfiguration. 
\figurename~\ref{fig:reconf-changes} shows an example illustrating this process. In the original configuration, we have two clusters with three clients each, and a new client $C_7$ joins the pipeline causing its reconfiguration. 

Depending on which configuration strategy we use (e.g., minimizing per-round communication cost and maximizing cluster data diversity~\cite{deng_hier}), the best-fit configuration for a given environment may vary. Let us consider an example of a best-fit configuration, shown on the right in \figurename~\ref{fig:reconf-changes}, to demonstrate how configuration differences affect the changes to be made when reconfiguring. In this new configuration, four existing nodes are assigned to a different LA, thus both clusters are affected. If we define $\Delta C$ as the set of reconfiguration changes, five changes take place to create the new configuration ($|\Delta C|=5$). Additionally, the cost per global aggregation round is likely to be different for the original and new configuration since new clusters are defined and nodes with different network distances are used as clients and aggregators in the pipeline.

\subsection{Reconfiguration cost}

In general, the reconfiguration cost \(\Psi_{rec}\) can be expressed with two values:

\begin{equation} \label{eq:rec-cost}
    \Psi_{rec} = (\Psi_{rc}, \Psi_{pr}),
\end{equation}
where \(\Psi_{rc}\) is the cost of reconfiguration changes and \(\Psi_{pr}\) is the post-reconfiguration cost defined as follows: 

\begin{equation} \label{eq:rc-cost}
    \Psi_{rc} = \sum_{i=1}^{|\Delta C|} \psi_{rc}(i), \; \Psi_{rc} \geq 0 \\
\end{equation}
\begin{equation} \label{eq:pr-cost}
    \Psi_{pr} = \Psi_{gr}^{new} - \Psi_{gr}^{orig} = \Delta \Psi_{gr}, \; \Psi_{pr} \in (-\infty, +\infty)
\end{equation}

Equation~(\ref{eq:rc-cost}) defines the cost of reconfiguration changes which is the sum of the cost for all $\Delta C$ reconfiguration changes. $\Psi_{rc}$ is a fixed cost incurred once during reconfiguration (expressed in cost units). It can be equal to or greater than zero, since a reconfiguration change either generates cost or has no associated cost (when a client fails or leaves the pipeline). Equation~(\ref{eq:pr-cost}) defines the post-reconfiguration cost $\Psi_{pr}$ which is the difference between the cost per global round of the new and the original configuration. Since $\Delta \Psi_{gr}$ is a recurring cost for each global round, it is given in cost units per round. $\Psi_{pr}$ can be negative, which means that the post-reconfiguration leads to a gain, i.e., the costs are reduced after reconfiguration. For example, if a reconfiguration removes a client from the pipeline, the communication cost is reduced since one model update is no longer sent to an LA.

In this paper, we focus on reconfiguration cost with respect to communication. Communication cost is usually defined as the traversed distance by packets~\cite{deng_hier} or as network bandwidth consumption on the path between communicating nodes~\cite{wang_hier}. The longer the distance a packet has to travel or the higher the bandwidth along the path of the packet, the higher the communication cost.
In HFL, this refers to the communication cost of transmitting model updates from clients to LAs or from LAs to the GA, both during and after reconfiguration. 

Each reconfiguration change incurs a cost of downloading the HFL service artifacts, e.g., service container image, if the service is not already present on the node, and the cost of downloading the model from a parent aggregator. We define $S_{svc}$ as the size of the HFL service artifacts in bits and $\mathcal{M}$ as the model size in bits, where $AS$ is the artifact server node, e.g., a container image repository, and $PA$ is the parent aggregator node. Then, we can calculate the communication cost of each reconfiguration change $\psi_{rc}^{comm}$ as follows:
\begin{equation} \label{eq:rc-comm-cost}
    \psi_{rc}^{comm}(i) = S_{svc} \times l(n_i,AS)  + \mathcal{M} \times l(n_i,PA),\hspace{1.5mm}i \in \Delta C,
\end{equation}
\noindent where \(l(x,y)\) is the link cost between nodes \textit{x} and \textit{y} (in units per bit), and $n_i$ is the node affected by change $i$. If the service is already downloaded on node $n_i$, we set $l(n_i,AS) = 0$.

The post-reconfiguration cost is a recurring one that depends on the communication cost within one global round. If we define $K$ as the total number of local clusters (i.e., local aggregators), $N_i$ as the number of clients in cluster $i$ where $c_{ij}$ is its $j-$th client, $L$ as the number of local rounds, and $GA$ as the global aggregator node, the communication cost for one global round $\Psi_{gr}^{comm}$ in HFL can be defined as:
\begin{equation} \label{eq:gr-cost}
    \Psi_{gr}^{comm} = \Psi_{ga}^{comm} + \Psi_{la}^{comm}
\end{equation}

\begin{equation} \label{eq:ga-cost}
    \Psi_{ga}^{comm} = \sum_{i=1}^{K}(l(LA_{i},GA) \times S_{mu})
\end{equation}

\begin{equation} \label{eq:la-cost}
    \Psi_{la}^{comm} = L \times \sum_{i=1}^{K} \sum_{j=1}^{N_i} (l(c_{ij}, LA_{i}) \times S_{mu}),
\end{equation}

\noindent where $\Psi_{ga}^{comm}$ is the communication cost of global aggregation and $\Psi_{la}^{comm}$ is the cost of local aggregations. We can observe that in each global round, we perform $L$ local aggregations from clients to their local aggregators (\ref{eq:la-cost}) and one global aggregation from local aggregators to the global aggregator (\ref{eq:ga-cost}). \(S_{mu}\) represents the model update size. In the rest of this paper, $S_{mu} = \mathcal{M}$. However, more compact model update representations by means of compression are also possible~\cite{sattler_comm}.

\subsection{Reconfiguration validation algorithm}

\begin{figure}[htbp]
    \centering
    \subfloat[\small Event flow]{\includegraphics[width=0.23\textwidth]{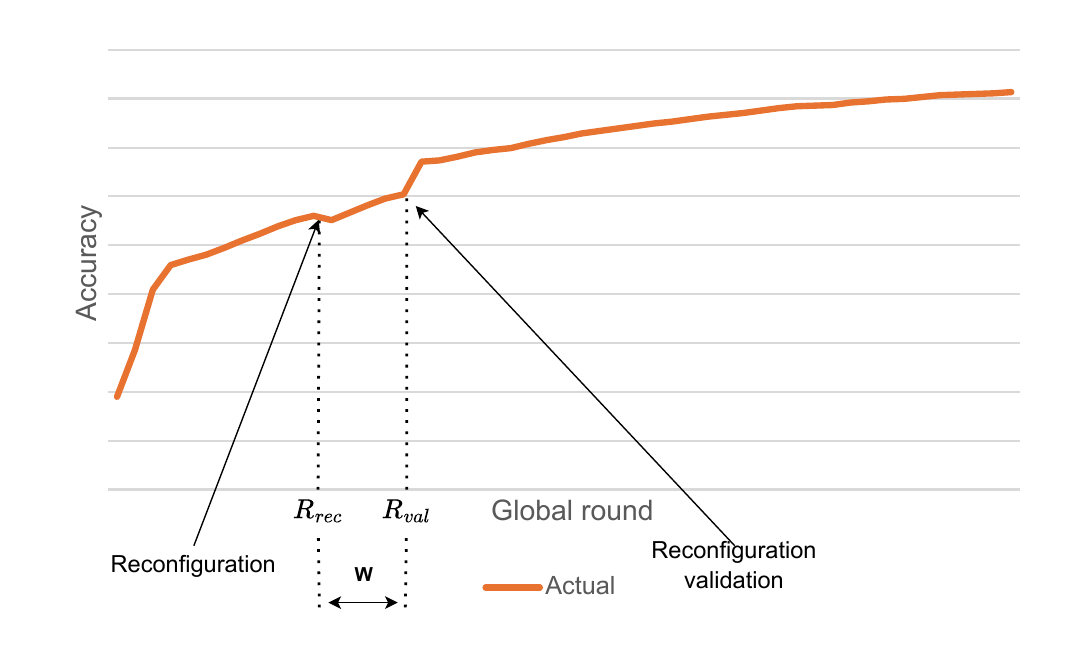}\label{fig:rva-flow}}
    \hfill
    \subfloat[\small Performance approximation]{\includegraphics[width=0.24\textwidth]{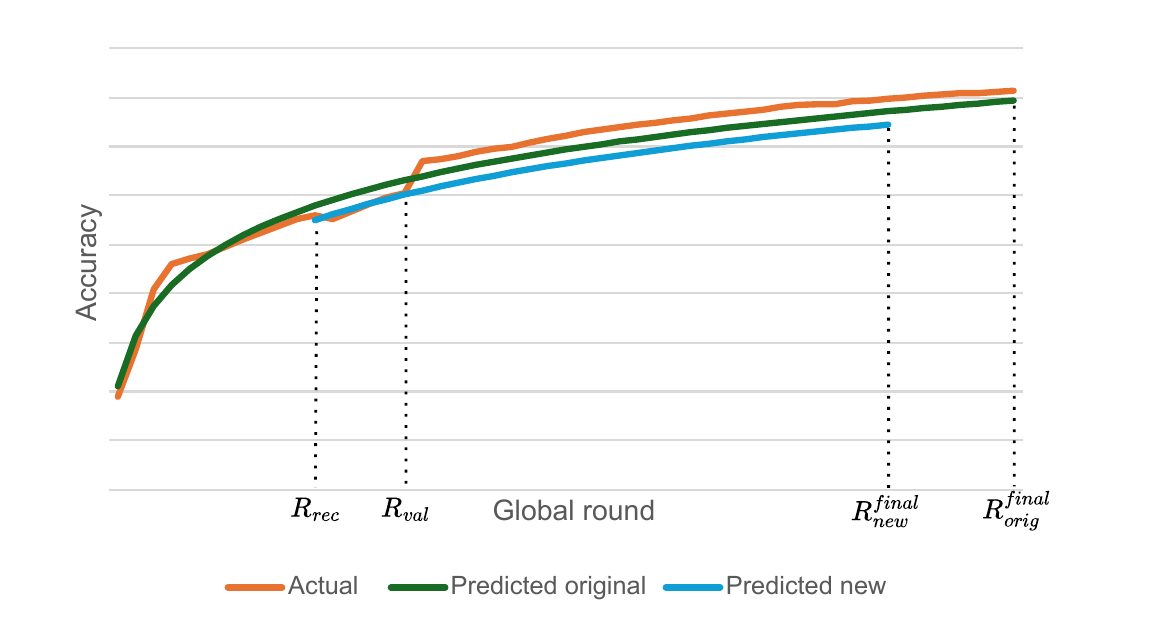}\label{fig:rva-approx}}
    \caption{RVA performing reconfiguration validation.}
    \label{fig:rva-example}
\end{figure}

Reconfiguration can sometimes have a negative impact on model performance or cost. To address this, we introduce a reactive reconfiguration validation approach \textit{Reconfiguration Validation Algorithm (RVA)} which is built into our orchestration framework. RVA validates the reconfiguration status after a specific number of global rounds post-reconfiguration, denoted as the validation window $W$. RVA can decide to keep the new configuration or to revert the pipeline to the original configuration. \figurename~\ref{fig:rva-flow} shows an example of the sequential flow of reconfiguration validation. An event occurs at round $R_{rec}$ and the reconfiguration is performed at the end of the current global round.\footnote{If the occurring event is a node leaving the system, the reconfiguration is postponed for at least $W$ rounds. The reason for this is that, compared to other events such as a new client joining or network changes, we do not know how the original configuration would behave without the missing node. Thus, we need to wait at least a few rounds to analyze the ML model behavior with the original configuration without the missing node.} After that, the orchestrator waits $W$ rounds until $R_{val}$ when it performs \textit{reconfiguration validation}. In this example, the validation decision is based on the observation that the last reconfiguration leads to a degradation of model performance, and the original configuration is restored. 

The decision whether the configuration leads to an improvement or degradation of the model performance is made based on the approximation of the future behavior of the original and new configuration. This approximation is made by regression based on the actual model performance values. An example of approximation functions is shown in \figurename~\ref{fig:rva-approx}. We can see that the prediction of the original configuration is made based on the observed performance before the reconfiguration, while the prediction of the pipeline performance with the new configuration is made based on the observed performance during the validation window. After calculating the regression functions for both configurations, the orchestrator predicts the performance value \emph{of the round in which the communication budget is exceeded}, $R_{final}$. The final round can be determined as follows: 
\begin{equation} \label{eq:r-fin}
    R_{final}^{orig} = R_{val} + \frac{B_{rem} - \Psi_{rc}}{\Psi_{gr}^{orig}}, \quad 
    R_{final}^{new} = R_{val} + \frac{B_{rem}}{\Psi_{gr}^{new}}
\end{equation}
\noindent where $R_{final}^{orig}$ and $R_{final}^{new}$ stand for the final rounds with the original and new configuration, respectively, and $B_{rem}$ is the remaining budget. The number of remaining rounds with the original configuration depends also on the cost of reconfiguration change, since additional cost is introduced when restoring the original configuration.

\begin{algorithm}[htbp]
\caption{Reconfiguration validation algorithm.}
\label{alg:rva}
\begin{algorithmic} [1]
{\ttfamily \scriptsize{
\Procedure{Reconfiguration}{$event, progress, task, orch$}
\State $origConfig \gets progress.config$
\State $newConfig \gets orch.calcBestFitConfig(orch.environment)$ \label{line:best-fit}
\State $\Psi_{rc} \gets getReconfChangeCost(origConfig, newConfig)$ 
    \Statex \hfill (eq. \ref{eq:rc-comm-cost}) \label{line:change-cost}
\If{$event.type = "nodeLeft"$}
    \State $waitNRounds(orch.W)$ \label{line:wait}
\EndIf
\State $R_{cur} \gets progress.currentRound$
\State $schedule(recVal(progress, origConfig, task.objective, R_{cur}),$
    \Statex \hfill $R_{cur} + orch.W)$
\State $progress.totalCost += \Psi_{rc}$
\State $deployConfiguration(newConfig)$
\EndProcedure

\Function{RecVal}{$progress, origConfig, objective, R_{rec}$}
\State $newConfig \gets progress.config$
\State $\Psi_{rc} \gets getReconfChangeCost(newConfig, origConfig)$ 
    \Statex \hfill (eq. \ref{eq:rc-comm-cost})
\State $\Psi_{gr}^{orig} \gets calcPerRoundCost(origConfig)$ (eq. \ref{eq:gr-cost}) \label{line:cpr-orig}
\State $\Psi_{gr}^{new} \gets calcPerRoundCost(newConfig)$ (eq. \ref{eq:gr-cost}) \label{line:cpr-new}
\State $f_{orig} \gets getPerformanceApproxFunction($
        \Statex \hfill $progress.accuracies[:R_{rec}])$ \label{line:approx-orig}
\State $f_{new} \gets getPerformanceApproxFunction($
        \Statex \hfill $progress.accuracies[R_{rec}:])$ \label{line:approx-new}
\State $R_{cur} \gets progress.currentRound$
\State $B_{rem} = objective.budget - progress.totalCost$
\State $R_{final}^{orig} \gets calcFinalRound(R_{cur}B_{rem},\Psi_{rc},\Psi_{gr}^{orig})$
    \Statex \hfill (eq. \ref{eq:r-fin})
\State $R_{final}^{new} \gets calcFinalRound(R_{cur},B_{rem},\Psi_{gr}^{new})$
    \Statex \hfill (eq. \ref{eq:r-fin})
\State $A_{final}^{orig} \gets f_{orig}(R_{final}^{orig})$
\State $A_{final}^{new} \gets f_{new}(R_{final}^{new})$
\If{$A_{final}^{orig} > A_{final}^{new}$} \label{line:bud-start}
    \State $progress.totalCost += \Psi_{rc}$
    \State $deployConfiguration(origConfig)$ \label{line:bud-end}
\EndIf
\EndFunction
}}
\end{algorithmic}
\end{algorithm}

Algorithm~\ref{alg:rva} presents the steps of the RVA. An infrastructure change event triggers the calculation of the new best-fit configuration (line \ref{line:best-fit}) using a selected configuration strategy. The orchestrator then calculates the cost of reconfiguration change to switch to the new configuration (line \ref{line:change-cost}), adds this cost to the total cost, and deploys the pipeline with the new configuration.\footnote{In the event of a node leaving the system, the orchestrator schedules a reconfiguration after $W$ global rounds (line \ref{line:wait}).} The orchestrator then schedules the reconfiguration validation after $W$ global rounds.

When the validation function is triggered, the orchestrator calculates the cost per global round for both the original (line \ref{line:cpr-orig}) and new configuration (line \ref{line:cpr-new}). The approximation functions are then determined by regression (lines \ref{line:approx-orig} and \ref{line:approx-new}). Note that for the approximation of the original function it only considers the values of accuracy up to the reconfiguration round, and for the new configuration it uses the values obtained during the validation window. Next, the orchestrator predicts the final accuracy value for both the original and new configuration. If the value of the original configuration is expected to be higher compared to the new configuration, it reverts the pipeline to the original configuration (lines \ref{line:bud-start} - \ref{line:bud-end}).

\section{Experimental evaluation}
\label{section:eval}

To evaluate our HFL orchestration framework and RVA, we conducted a series of testbed experiments on a K3s~\cite{k3s} cluster, a lightweight distribution of Kubernetes tailored to edge settings. We deployed a cluster of 13 nodes: one controller and 12 worker nodes. All nodes were VMs running \textit{Ubuntu Server 22.04} with 2 CPU cores and 4 GB of RAM.

\begin{figure}[htbp]
\centering
\includegraphics[width=3.4in]{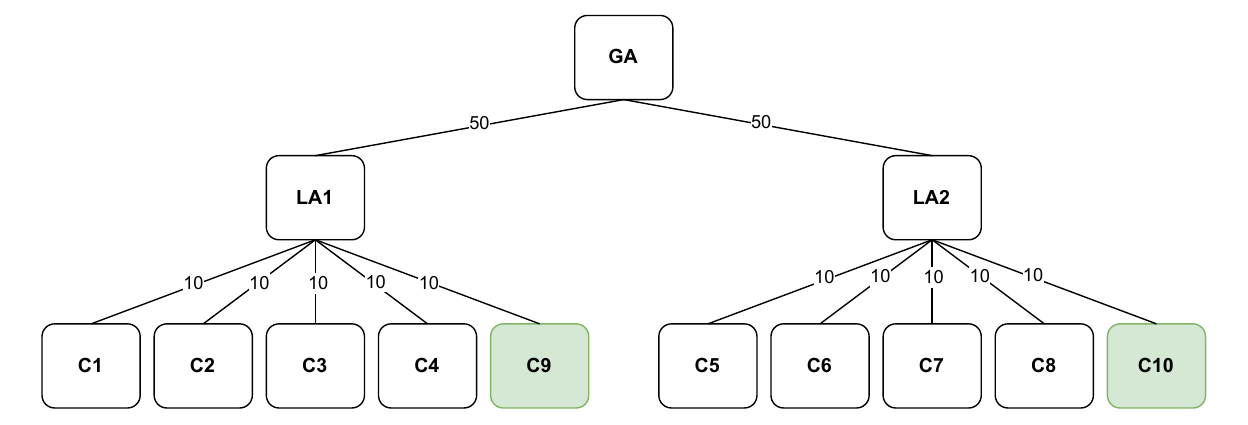}
\caption{Experimental evaluation: topology.}
\label{fig:exp-setup}
\vspace{-2mm}
\end{figure}

In all experiments, we deploy the nodes based on the infrastructure topology shown in \figurename~\ref{fig:exp-setup}. It depicts the nodes placed in clusters and the communication cost between a node and its parent, defined in \textit{units per MB}. At the beginning of each experiment, two clusters are deployed consisting of four clients with two LA nodes ($LA_1$ and $LA_2$) and the top $GA$ node. After 10 rounds ($R_{rec}$), two new client nodes $C_9$ and $C_{10}$ are added to the topology using the configuration strategy that minimizes communication cost between clients and their LAs (an adaptation of~\cite{deng_hier}). All configuration parameters are specified in Table \ref{table:exp-par}.
\begin{table}[h!]
\centering
\caption{Experimental evaluation: configuration parameters.}
\label{table:exp-par}
\begin{tabular}{ c | c}   
 \hline
 Communication Budget & 100,000 units \\
 Configuration Strategy & \textit{minCommCost} \\
 Local Epochs & 2 \\
 Local Rounds & 2 \\
 $S_{mu}$ & 3.3 MB \\
 $R_{rec}$ & 10 \\
 $W$ & 5 \\
 Regression Function & Logarithmic \\
 \hline
\end{tabular}
\vspace{-4mm}
\end{table}
Clients train a CNN model for an image classification task on the CIFAR-10 dataset in all experiments. 
The CNN model used has two convolutional layers (6 and 16 channels) each followed by a ReLU activation function and a 2x2 max pooling. After flattening the output of the last layer, the model includes two fully connected layers with 120 and 84 units respectively, both using ReLU activation functions.
The output layer of 10 units, for each class in CIFAR-10.

The dataset on each client was specified for each experiment separately to show the difference in performance with different data distributions. 
We evaluate RVA with two types of data distributions on the client nodes: IID and non-IID. In the IID setup, all clients bring in similar datasets containing all ten classes of the CIFAR-10 dataset and an equal number of samples per class. In the non-IID setup, each client has two classes in its dataset and the number of samples per class is equal. Table \ref{table:data-distr} shows the data distribution per client for each scenario evaluating RVA. 

\begin{table}[h!]
\centering
\caption{Client data distribution per evaluation scenario.}
\label{table:data-distr}
\begin{tabular}{ c | c | c | c | c }   
    \hline
    \textbf{Client} & \textit{1.a} (I) & \textit{1.b} (I) & \textit{2.a} (NI) & \textit{2.b} (NI) \\
    \hline
    $C_1$ & S & S & [0, 1] & [0, 1] \\
    $C_2$ & S & S & [2, 3] & [2, 3] \\
    $C_3$ & S & S & [4, 5] & [4, 5] \\
    $C_4$ & S & S & [6, 7] & [6, 7] \\
    $C_5$ & S & S & [0, 1] & [0, 1] \\
    $C_6$ & S & S & [2, 3] & [2, 3] \\
    $C_7$ & S & S & [4, 5] & [4, 5] \\
    $C_8$ & S & S & [6, 7] & [6, 7] \\
    \rowcolor{SeaGreen3!30!} $C_9$ & S & L & [0, 1] & [8, 9] \\
    \rowcolor{SeaGreen3!30!} $C_{10}$ & S & L & [0, 1] & [8, 9] \\
    \hline
\end{tabular}
\newline
\newline
\footnotesize{I - IID; NI - Non-IID} \\
\footnotesize{S - small IID dataset (100 samples per class)} \\
\footnotesize{L - large IID dataset (1000 samples per class)} \\
\footnotesize{[0, 1] - classes \textit{0} and \textit{1} (1000 samples per class)}
\end{table}

Two scenarios were used to evaluate RVA: scenario \textit{1} with IID data and scenario \textit{2} with non-IID data. 
Each scenario consists of two sub-scenarios that are designed to show examples when new nodes either improve (\textit{1.b} and \textit{2.b}) or degrade (\textit{1.a} and \textit{2.a}) model performance. 
Degraded performance means that the final model accuracy with the new configuration would be lower than with the original one.
In scenario \textit{1.a} the model performance is not expected to improve much as new clients bring in comparable datasets to those already present in the setup, while in scenario \textit{1.b} new clients introduce larger datasets, which is expected to significantly improve the model. Similarly, in scenario \textit{2.a} new clients bring in classes that are already covered within clusters which is not expected to impact model performance, while in \textit{2.b} clients bring in two missing classes in both clusters.

\begin{figure}[htbp]
\centering
\includegraphics[width=3in]{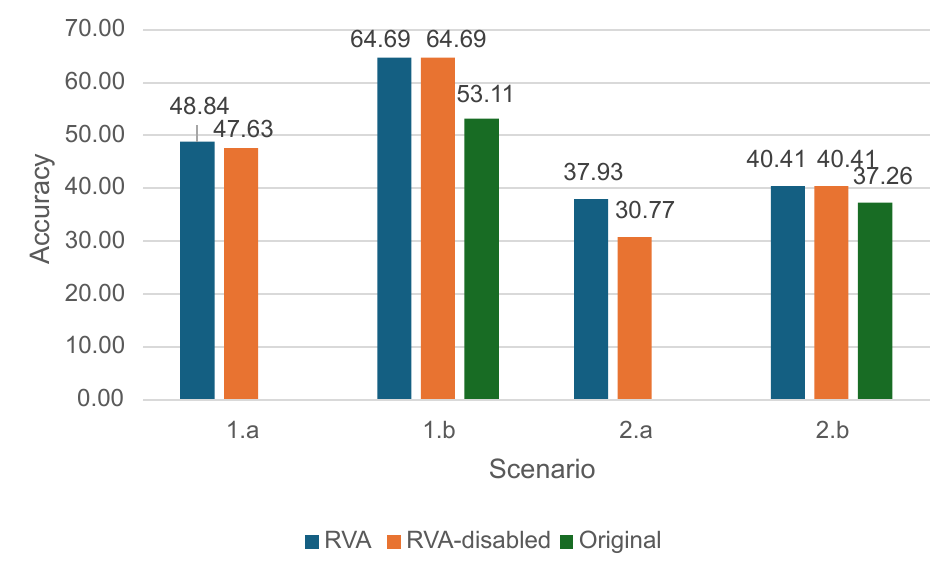}
\caption{RVA evaluation: results.}
\label{fig:res-acc}
\vspace{-2mm}
\end{figure}

The model accuracy observed in the end of all scenarios are shown in \figurename~\ref{fig:res-acc}. When RVA is used to revert the pipeline to the original configuration in which new nodes degrade the performance (\textit{1.a} and \textit{2.a}), we note a consistent improvement of the model accuracy under a limited communication budget, compared to the setup that is not using RVA (\textit{RVA-disabled}). 
The results also show that RVA decided correctly to keep the new pipeline configuration in scenarios \textit{1.b} and \textit{2.b}~\footnote{The result of RVA is thus the same as with RVA-disabled.} by comparing the final accuracy of the setup which uses RVA with a setup that erroneously decided to revert the pipeline to the original configuration (\textit{Original}). These results show that RVA detects events that introduce both improvement and degradation of model performance and makes correct decisions favoring performance under a limited communication budget.

\begin{figure}[htbp]
    \centering
    \subfloat[\small Accuracy]{\includegraphics[width=0.24\textwidth]{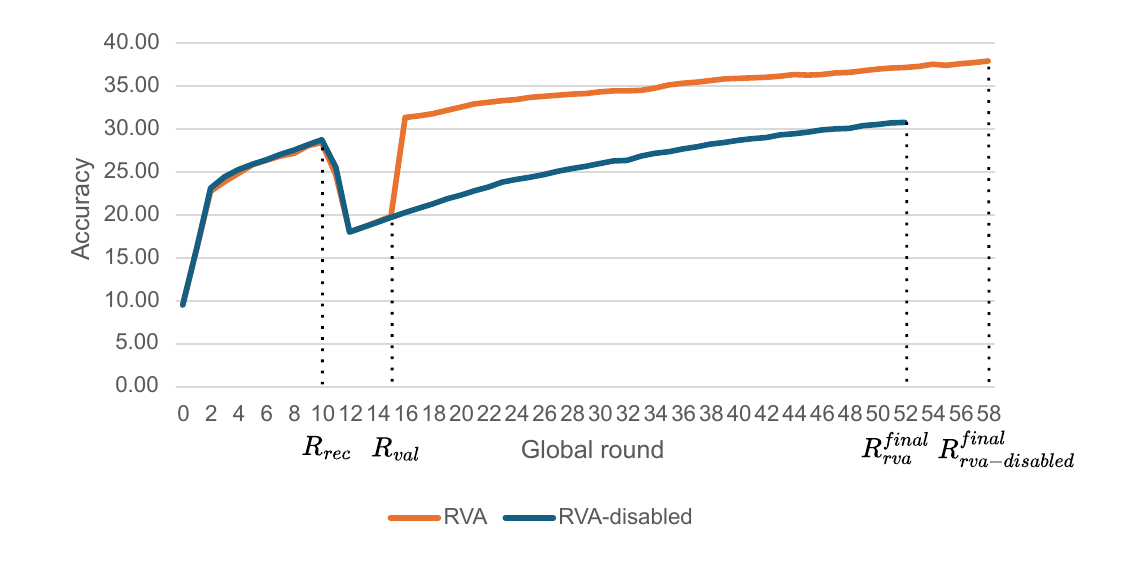}\label{fig:non-deg-acc}}
    \hfill
    \subfloat[\small Total cost]{\includegraphics[width=0.24\textwidth]{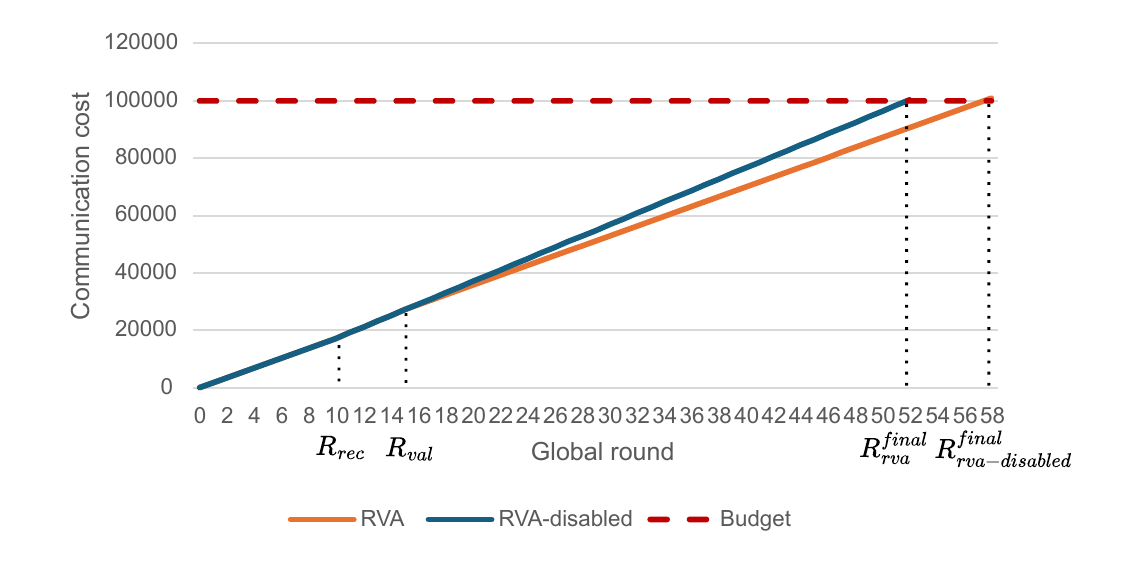}\label{fig:non-deg-cost}}
    \caption{Scenario \textit{2.a}: trend of accuracy and cost over global rounds.}
    \label{fig:non-deg-res}
\end{figure}

In \figurename~\ref{fig:non-deg-res} we analyze how accuracy and cost change over global rounds for scenario \textit{2.a} with non-IID datasets. We can observe a significant drop in model accuracy (\figurename~\ref{fig:non-deg-acc}) after reconfiguration in round $R_{rec}$. As the new configuration introduced higher cost per round, the budget without RVA is exceeded earlier (\figurename~\ref{fig:non-deg-cost}) while reaching lower accuracy compared to the configuration that used RVA. Thus, by validating reconfiguration decisions at runtime, RVA enables the selection of configurations that achieve improved model performance within a communication cost budget.   

Note that our framework introduces minimal computing overhead to the system (15\,MB RAM, 0.15 cores). Also, it needs significantly more time to detect a new node (15\,s) than node removal (0.5\,s) due to K3s limitations. Prompt reaction to node removal is more important in our use cases, as the orchestrator can quickly reconfigure the pipeline with the best-fit configuration without the removed node.

\section{Related work}
\label{section:rel}

A significant number of works~\cite{Lin24DFL, deng_hier, Trindade24, Xu23, Pervej24, Heydar20,lackinger2024inference,zhang2024optimizing} have studied different variants of the HFL pipeline configuration problem, addressing aspects such as topology formation (client-LA association) and configuring how aggregation and model distribution take place. These works do not address HFL pipeline reconfiguration or offer limited (periodic) reconfiguration support. 

Dyn-FedCH~\cite{wang2022accelerating} and HiFlash~\cite{Wu23hiflash} focus on environments where global aggregation takes place asynchronously, which introduces the problem of stale updates. Dyn-FedCH and, similarly, DCFL~\cite{wang_hier}, define weights that encode the importance of stale updates during global aggregation, while HiFlash proposes a reinforcement learning (RL) scheme that dynamically adapts these weights. Topology reconfiguration takes place when an LA is (de)activated (HiFlash) or when many LAs include straggler clients (Dyn-FedCH, DCFL). These approaches \emph{assume different HFL architectures and do not evaluate the impact of topology reconfiguration decisions}, which is hard to predict in advance. Sai et al.~\cite{Yinghui24} also assign aggregation weights to each client and use RL to adapt them at every global round. The reward received at each adaptation decision accounts for both accuracy improvement and cost reduction, and per-round topology is implicitly configured by only including clients whose weight exceeds a threshold; \emph{contrary to our scheme and many other works in the literature, client-LA association is fixed}. HED-FL~\cite{Rango23} generalizes to multi-level FL hierarchies, where the frequency of communication rounds is adapted based on the observed accuracy. In contrast with our target scenarios, \emph{a fixed tree topology is assumed} where at each round an aggregator randomly selects clients or aggregators from the level below.

Notably, the aforementioned works \emph{do not deal with orchestrator design aspects}, often neglecting significant practical reconfiguration costs. Let us stress that different device-LA association algorithms from the literature can be integrated into our orchestration framework, without affecting the functionality of our RVA.

\section{Conclusion and future work}

Many prior studies have explored various aspects of HFL pipeline configuration but often overlook runtime reconfiguration and its impact on ML model performance. This paper introduces a framework for adaptive orchestration of HFL pipelines, enabling runtime reconfigurations while estimating their costs and predicting their impact on future ML performance. The framework incorporates mechanisms to reactively validate reconfiguration decisions using the proposed RVA and, if necessary, revert to the original configuration. Our evaluation in a realistic K3s cluster demonstrates that reconfiguration validation during runtime can improve ML model accuracy within the limited communication cost budget.
Future work will address current limitations in our scheme's evaluation, particularly focusing on two aspects: (i) measuring RVA performance under more complex datasets and tasks, and (ii) demonstrating the ability of our framework to support additional orchestration objectives, such as completion time or energy consumption minimization. Finally, our plan is to extend RVA's predictive capabilities by also considering node utility.

\section*{Acknowledgments}
This work has been supported by the European Union’s Horizon Europe research and innovation programme under grant agreement No. 101079214 (AIoTwin), and by the Croatian Science Foundation under the project DOK-2020-01-1430.

\bibliographystyle{IEEEtran}
\bibliography{IEEEabrv,references.bib}

\end{document}